\documentclass[useAMS,usenatbib]{mn2e}
\usepackage{subfigure}
\usepackage{lscape}
\usepackage[dvips]{color}
\usepackage{amssymb}
\usepackage{graphicx}
\usepackage{ulem} 
\newcommand{\apj}{ApJ}

\newcommand{\aj}{AJ}

\newcommand{\aap}{A\&A}

\newcommand{\pasp}{PASP}

\title[Detection of globular clusters]{The detection of globular clusters in galaxies as a data mining problem}

\author[Brescia M. et al. 2011]{Massimo Brescia$^{1}$, Stefano Cavuoti$^{2}$, Maurizio Paolillo$^{2}$, Giuseppe Longo$^{2,3}$,
\and Thomas Puzia$^{4}$ \\ \\ 
1 - INAF-Astronomical Observatory of Naples, via Moiariello 16, I-80131 Napoli, Italy\\
2 - Dipartimento di Scienze Fisiche, University Federico II, via Cinthia 6, I-80126 Napoli, Italy\\ 
3 - Visiting Associate, California Institute of Technology, Pasadena USA \\
4 - Department of Astronomy and Astrophysics, Pontificia Universidad Cat—lica de Chile, Santiago, Chile\\} 

\begin{document}

\date{Accepted ;
  Received ; in original form }
\pagerange{\pageref{firstpage}--\pageref{lastpage}} \pubyear{2011}

\maketitle
\label{firstpage}

\begin{abstract}
We present an application of self-adaptive supervised learning classifiers derived from the Machine Learning 
paradigm, to the identification of candidate Globular Clusters in deep, wide-field, single band HST images. 
Several methods provided by the DAME (Data Mining \& Exploration) web application, were tested and compared 
on the NGC1399 HST data described in \citet{paolillo11}. 
The best results were obtained using a Multi Layer Perceptron with Quasi Newton learning rule which achieved 
a classification accuracy of $98.3 \%$, with a completeness of $97.8\%$ and $1.6\%$ contamination. An extensive set of experiments revealed that 
the use of accurate structural parameters (effective radius, central surface brightness) does improve the final result, but only by $\sim 5 \%$.
It is also shown that the method is capable to retrieve also extreme sources (for instance, very extended 
objects) which are missed by more traditional approaches.
\end{abstract}


\section{INTRODUCTION}
\label{intro}

The need to effectively exploit the scientific information contained in current and future synoptic surveys 
has led to a renaissance of interest in the application of Data Mining (DM) methods to astronomical programs. 
DM, in fact, seems among the few, if not the only, ways to cope with the complexity and size of existing 
and foreseen massive data sets such as, for instance, those expected to be provided by the LSST. 
The DM methods, however, are also very useful to capture the complexity of small 
data sets and, therefore, can be effectively used to tackle problems of much smaller scale. 
In this paper we used a variety of methods provided by the DAta Mining \& Exploration Web Application 
REsource (DAMEWARE, \emph{http://dame.dsf.unina.it/beta\_info.html}) for the identification of Globular Clusters (GCs) in the galaxy NGC1399 using single band photometric 
data obtained with the Hubble Space Telescope (HST). 

The identification and physical characterization of Globular Cluster (GC) populations in external galaxies is 
of interest to many astrophysical fields: from cosmology, to the evolution of star clusters and galaxies, to the formation 
and evolution of binary systems. 
The identification of Globular Clusters in external galaxies usually requires the use of wide-field, multi-band 
photometry since in galaxies located more than a few Mpc away they appear as unresolved sources 
in ground-based astronomical images and are thus hardly distinguishable from background galaxies which introduce 
significant contamination problems. 
For such reason, GCs are traditionally selected using methods based on their colors and luminosities. 
However, in order to minimize contamination and to measure GC properties such as sizes and structural parameters 
(core radius, concentration, etc.), high-resolution data are required as well which,  for star clusters outside the Local Group, are 
available only through the use of space facilities (i.e. Hubble Space Telescope, HST).
Obtaining suitable HST data is however challenging in terms of observing time since the optimal datasets should 
be: i) deep, in order to sample the majority of the GC population and ensure the high S/N required to measure 
structural parameters \citep[see e.g.][]{car001}; ii) with wide-field coverage, in order to minimize projection 
effects as well as to study the overall properties of the GC populations, which often differ from those inferred 
from observations of the central region of a galaxy only; and iii) multi-band, to effectively select GC based on 
colors.

It is apparent that, in order to reduce observing costs, it would be much more effective to use single-band HST 
data.
Such approach however requires to carefully select the candidate GCs based on the available photometric and 
morphological parameters in order to avoid introducing biases in the final sample (see below). 
Here we intend to show that the use of properly tuned DM algorithms can yield very complete datasets with low 
contamination even with single band photometry, thus minimizing the observing time requirements and allowing to 
extend such studies to larger areas and to the outskirts of nearby galaxies.

The paper is structured as follows: in Sect. \ref{thedata} we describe the data used to test of the various 
method; in Sect. \ref{DAMEWARE} we provide a short methodological and technical introduction to DAMEWARE and to 
some classification methods tested for the first time in an astronomical context. 
In Sect. \ref{exp} and \ref{disc} we describe the results of the experiments and draw our conclusions.

\section{The data}\label{thedata}
The dataset used in this experiment consists of wide field HST observations of the giant elliptical NGC1399 located
at the heart of the Fornax cluster. 
This galaxy represents an ideal test case since, due to its distance \citep[$D=20.13\pm0.4$ Mpc, see][]{dunn06}, it is possible to cover a large fraction 
of its GC system (out to $>5 R_e$) with a limited number of observations. 
Furthermore, it is particularly challenging because, at this distance, GCs are only marginally resolved even by HST; in fact at NGC~1399 distance 1 ACS pixel corresponds to 2.93 pc (1\arcsec\ = 97.7 pc). 
This dataset was used by \citet{paolillo11} to study the GC-LMXB connection and by Puzia et al. (2012, in preparation) to study the structural properties of the GC population. 
We summarize below the main properties of the dataset, and refer to these works for a more detailed description of the observations and of data analysis.

\begin{figure}
 \includegraphics[width=8cm]{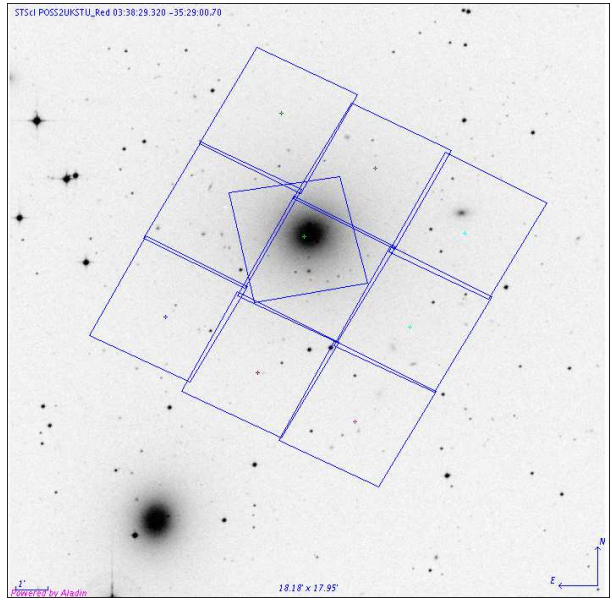}
 \caption{The field of view covered by the 3x3 HST/ACS mosaic in the F606W band. The central field, with a different orientation, shows the region covered by previous archival ASC observations in $g$ and $z$ bands. }
 \label{FOV}
\end{figure}

\begin{figure}
\includegraphics[width=8.5cm]{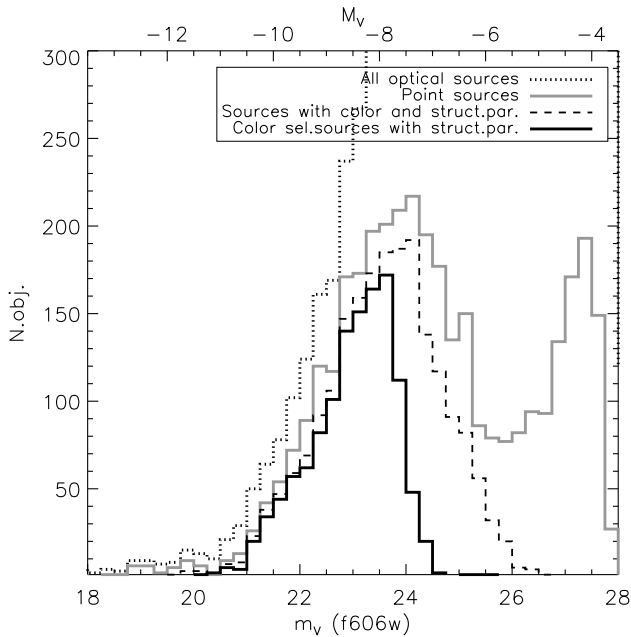}
\caption{Luminosity distributions of all detected (dotted line) and point-like (e.g. with stellarity index $>0.9$, solid gray line) sources§ within the HST FOV. Also shown are the two additional subsamples discussed in \S \ref{thedata} and \S \ref{exp}: the KB composed of sources with both color and structural parameters (dashed line), and the subset of bona-fide color-selected GCs based on Table \ref{col_sel} (solid black line).}
\label{maghisto} 
\end{figure}

The optical data were taken with the HST Advanced Camera for Surveys (ACS, program GO-10129, PI T.Puzia), in the 
F606W (broad $V$ band) filter, with integration time of 2108 seconds for each field. The observations were arranged in a 3x3 ACS mosaic, 
and combined into a single image using the MultiDrizzle routine \citep{koekemoer02}. 
The final scale of the images is 0.03\arcsec/pix, providing Nyquist sampling of the ACS PSF. 
The field of view of the ACS mosaic covers ~100 square arcmin (Figure \ref{FOV}) extending out to a projected galactocentric 
distance of $\sim 55$ kpc, i.e. $4.9 r_e$ of the GC system ($\sim 5.7 r_e^{gal}$). 
The source catalog was generated with \textsc{SExtractor} by imposing a minimum area of 20 pixels: it contains 12915 sources and 
reaches $7\sigma$ detection at $m_V=27.5$, i.e. 4 mag below the GC luminosity function turnover, thus allowing to sample the 
entire GC population (see Figure \ref{maghisto}). 
The catalog astrometric solution was registered to the USNO-B1 reference frame, obtaining a final accuracy of 0.2" r.m.s. 

For 4239 sources we were able to measure structural parameters (which require very high S/N, see \citealt{car001} and Puzia et al. 2012), fitting King surface brightness profile models with the \textsc{Galfit} software \citep{peng02}, and deriving tidal, core, effective radii 
and central surface brightness values for each cluster. The accuracy of these measurements was estimated simulating several thousand artificial GCs with the \textsc{Multiking} code (available at: \emph{http://people.na.infn.it/paolillo/Software.html}) specifically written to account for ACS field distortion, PSF variation, dithering pattern \citep[Puzia et al. 2012]{paolillo11}.

The NGC1399 region covered by our mosaic lacks color informations for all HST F606W sources. 
In this paper we shall therefore make use of two ancillary multiwavelength datasets: archival HST $g-z$ observations \citep{kundu05}, which cover the very central region of the galaxy ($10\%$ of the sample, see Figure \ref{FOV}), and 
$C-T1$ ground based photometry from \cite{bassino06}, covering the whole mosaic. 
The latter is only available for $\sim  14\%$ of our sources, and due to background light contamination it is very 
incomplete in the proximity of the galaxy center. In total 2740 sources of the catalog have multi-band (either $g-z$ or $C-T1$) photometry.

Finally, the subsample of sources used to build our Knowledge Base (KB, see \S\ref{DM_sec}) to train the DM algorithms, is composed by the 2100 sources with all photometric and morphological informations: isophotal magnitude, kron radius, aperture magnitudes within a 2, 6 and 20 pixels (corresponding to $0.06^{\prime\prime}$, $0.18^{\prime\prime}$, and $0.6^{\prime\prime}$) diameter, ellipticity, position angle, FWHM, \textsc{SExtractor} stellarity index, King's tidal and core radii, effective radii, central surface brightness, and either $g-z$ or $C-T1$ color. The magnitude distribution of such subsample is shown in Figure \ref{maghisto} as a dashed line.

The typical choice to select GCs based on multi-band photometry would be to adopt the magnitude and color cuts 
reported in Table \ref{col_sel}, and highlighted in Figure \ref{col_mag} with a dashed line; the magnitude limit $z<22.5$ does not exploit the full depth of the HST data but is adopted to be consistent with the $T1<23$ limit used for the ground-based colors, thus ensuring a uniform limit across the whole field-of-view. In the following we thus assume that bona-fide GCs are represented by such sources, in order to explore how well different selection methods based on single band photometry are able to retrieve the correct population of objects. The F606W magnitude distribution of color-selected GCs is shown in Figure \ref{maghisto} as a black solid line.

\begin{table}
\caption{Photometric selection criteria for GC candidates}
\label{col_sel}
\begin{tabular}{lcc}
\hline
 & color cut & magnitude cut\\
\hline
Ground-based & $1.0\leq C\!-\!T1<2.2$ & $T1<23$\\
data &  & \\\\
HST data & $1.3\leq g\!-\!z<2.5$ & $z<22.5$ \\
\hline
\end{tabular}
\end{table}

\section{Some considerations on Data Mining}\label{DAMEWARE}
\label{DM_sec}
DAMEWARE \citep{brescia2010} is one of the main products made available through the DAME (Data Mining \& Exploration) 
Program Collaboration.
It provides a web browser based front-end, able to configure data mining experiments on massive data sets and to execute  
them on a distributed computing infrastructure (cloud/grid hybrid platform).
DAMEWARE offers the possibility to access different DM functionalities (supervised classification, regression and 
clustering) implemented with different methods (traditional MLPs, Support Vector Machines, etc.).
Even though specifically designed to deal with massive data sets, DAMEWARE can also be used on small ones. 
It needs however to be taken into account that, due to the poor coverage of the parameter space by the KB, DM on small data 
sets requires special care. 
In what follows we shall outline the main strategy behind our procedure.

The problem tackled in this work is a typical supervised classification task and therefore, while referring
the reader to \citet{duda} and \citet{bishop1995} for a general introduction to DM, we shall shortly summarise 
some aspects which are relevant to the experiments described in the next paragraph. 

First of all, it needs to be kept in mind that in the data mining practice, there is no way to a priori 
select the algorithm which offers the best performances for a given task and that therefore a number of 
trial-and-error experiments must be performed in order to identify the method with the best performances. 
From a logical point of view, effective supervised classification is based on the following steps:

\begin{enumerate}
\item To select and create the data parameter space, i.e. to create the data input patterns (or features) 
to be submitted to the classifiers. It is important in this phase to build homogeneous patterns, i.e. with 
each pattern having the same type and number of parameters;
\item to prepare the datasets which are needed for the different experiment steps: training, validation and 
test sets (the dataset must include also target values for each input pattern, i.e. the desired output values, 
coming from any available knowledge source), by splitting the KB into variable subsets to be submitted at each 
phase; 
\item to analyze and select classification model, based on theoretical principles and on the user experience about 
the content of the KB;
\item to perform complete sequences of experiments with all model candidates and compare their results in terms of 
training error, learning robustness, output correctness (as defined below); this phase might also require a pruning of the parameter 
space);
\item finally, to identify the best model which will then be adopted as the final classifier to be applied to the 
entire dataset.
\end{enumerate}

\noindent Optionally (either because some methods do not require it or simply as an user choice) a validation 
procedure may be introduced.
Validation is the process of checking whether the classifier meets some criterion of generality when dealing 
with unseen data in order to avoid over-fitting or to stop the training on the base of an ''objective'' 
criterion. 
Here ''objective'' implies a criterion which is not based on the same data used for the training procedure. 
Obviously, validation requires an additional data set which can be prepared by the user directly or in an 
automatic fashion.

When the training set is of limited size - such as the one used in this paper - it is almost unavoidable to adopt 
a ''subset validation'' procedure. 
This implies the partitioning of a sample of data into subsets, such that the analysis is initially performed on a 
single subset, while the other subset(s) are retained for subsequent use in confirming and validating the initial 
analysis. 
In practice, the data sample is divided into $N$ subsets, some of which are used for the training phase (training set), 
while the others are employed, as validation sets, to compare the model prediction capability. 
By varying the value of $N$ (different splitting of the data sets) it is possible to evaluate the prediction accuracy 
of the trained model \citep{kotia}.

The so called $K$-fold cross-validation divides the whole dataset into $K$ subsets, each of them is alternately 
excluded from the validation set.
In practice all data are used for the training and test phases in an independent way. In this case we obtain $K$ 
classifiers $\left( 2 \leq K \leq  n \right)$ whose outputs can be used to obtain a mean evaluation. 
The downside of this method is that it is very expensive in terms of computing time in the case of massive datasets.

As it was briefly mentioned, in a supervised machine learning scheme, the training is done by means of a mechanism in which the model output is compared with the desired target output for each input pattern, allowing to define a training error. The choice of the metric function used for the comparison (which defines the training error) determines the evaluation criteria and the learning rule of the model.
Different error evaluation metrics exist in literature, depending on the problem complexity to be solved. 
In our experiments we used several methods.
The most common metric is the MSE (Mean Square Error) of the difference between model and target outputs.
Supervised neural networks that use MSE cost function can use formal statistical methods to determine 
the confidence of the trained model \citep{yang91} while the MSE computed on a validation set can be used as 
an estimate of the variance. 
This value can then be used to calculate the confidence interval of the output of the network assuming a normal 
distribution. 
A confidence analysis made in this way is statistically significant as long as the output probability distribution 
remains the same and the network is not modified.

By assigning a \textit{softmax} activation function \citep{bishop1995} on the output layer of the neural network (or a \textit{softmax} 
component in a component-based neural network) for categorical target variables, the outputs can be interpreted as 
posterior probabilities \citep{sutton1998}.
This is very useful in classification as it gives a certainty measure on classifications.

Many supervised models also support the use of the Cross Entropy error function for addressing classification problems 
in a consistent statistical fashion \citep{rubin2004}.

\noindent The Cross Entropy method consists of two phases:\\
\noindent 1- Generate a random data sample (trajectories, vectors, etc.) according to a specified mechanism;\\
\noindent 2-	Update the parameters of the random mechanism based on the data to produce a ''better'' sample in the next 
iteration. \\

\noindent In practice a data model is created based on the training set, and its cross-entropy is measured on a test set to assess how accurate the model is in predicting the test data.
In practice, the method compares two probability distributions, $p$ the true distribution of data in any corpus, and $q$ which is the distribution of data as predicted by the model.
Since the true distribution is unknown, cross-entropy cannot be directly calculated and, an estimate of cross-entropy is 
calculated using the following formula:

$$H\left(T,q \right)=- \sum_{i=1}^{N} \frac{1}{N} log_2 q\left(x_i \right)$$
where $T$ is the chosen training set, corresponding to the above mentioned true distribution $p$, $N$ is the number of objects in the test set, and $q\left( x \right)$ is the probability of the event $x$ estimated from the 
training set. 

Due to the supervised nature of the classification task, the system performance can be measured by means of a test
set during the testing procedure, in which unseen data are given to the system to be labelled. The overall performance thus integrates information about the classification accuracy (i.e. in terms of output correctness). Moreover, the results obtained from the unseen data are also important to evaluate the learning robustness, i.e. the generalization capability of the network in presence of data samples never used during the training phase.
However, when a data set is unbalanced (i.e. when the number of samples in different classes varies greatly) the error 
rate of a classifier is not representative of the true performance of the classifier itself. 

For the specific problem addressed in this paper we used five among the different classification methods available in
DAMEWARE. 
Namely: MLP-BP (Multi Layer Perceptron trained by Back Propagation), SVM (Support Vector Machines), GAME (Genetic 
Algorithm Model Experiment), MLPGA (MLP with Genetic Algorithms), and MLPQNA (Multi Layer Perceptron trained by Quasi 
Newton).
MLP-BP and SVM have already been described several times in the astronomical literature and therefore we refer the reader to \citet{bishop1995} and \citet{chang2011}. 
For what the other methods are concerned, since they are used for the first time in an astronomical context, we shall provide some further details.\\

\subsection{The Multi Layer Perceptron trained by Genetic Algorithms}
Genetic Algorithms (GA) are computational methods inspired to Darwin's evolutionary mechanism
\citep{holland75}. 
GA are particularly powerful in solving problems where the solution space is not well defined.
When they are embedded into a MLP network, the resulting learning algorithm (named MLPGA model) consists mainly in the 
cyclic exploration of the parameter space aimed at discovering the best solution \citep{mlpga}. 

\noindent In a Genetic Algorithm each element of a population (i.e. each data point) is called chromosome and 
is composed by a set of genes (features) that represents its DNA. From a more traditional point of view, each DNA can be
therefore considered as a possible solution to the problem.
The starting point of the method consists in the random generation of a population of chromosomes, 
for example by using normal or uniform statistical distributions. 
Then the method proceeds by cyclic variation and combination of the initial population, 
modifying their DNA's (neuron weights) according the standard feed-forward MLP calculations on input patterns.
The final goal is to find the best population (best problem solution) where ``best'' is defined according to some
fitness criterium.

In other words, at each evolutionary step (backward phase of the MLPGA model), 
the output chromosomes are obtained by applying several genetic operators to the input population 
and by evaluating through a specific fitness function the goodness of the newly generated population.
The fitness function provides a method to discard the worst chromosomes from the population thus allowing only the best
candidates to evolve to the next generation (similarly to what happens in natural selection).
The entire cycle is iterated until the chromosome with the desired fitness is found (i.e. the best solution to the classification problem).
The training error calculation follows the MSE criterion.

\subsection{The Genetic Algorithm Model Experiment}
As it was briefly mentioned above, this machine learning model arises from an original customization, made by DAME 
group, of the standard generalized GA model. 
All basic theoretical aspects for a generic GA have already been presented in the MLPGA section. 
The idea behind the Genetic Algorithm Model Experiment (GAME model) is to create a special fitness function, based 
on a polynomial expansion approximation, able to perform supervised adaptive learning on MDS. 
The analytical expression used to solve classification problem is the trigonometric series expansion of each input 
pattern features, compared with the corresponding known pattern target value. 
Then the whole error (MSE, Mean Square Error), which is the fitness function, is calculated at each cycle for 
all input patterns and the population of genetic chromosomes is updated according the classical genetic operators 
(crossover and mutation). 
This loop ends when the minimum error is found (below a chosen error threshold) or if 
the maximum number of iteration is reached.

\subsection{The Multi Layer Perceptron trained by Quasi Newton rule}
Quasi-Newton Algorithms (QNA) are variable metric methods for finding local maxima and minima of functions \citep{davidon}. 
The model based on this learning rule and on the MLP network topology is then called MLPQNA.
QNA are based on Newton's method to find the stationary (i.e. the zero gradient) point of a function. 
Newton's method assumes that the function can be considered as quadratic in a narrow region around the 
optimum and uses the first and second derivatives (gradient and Hessian) to find the stationary 
point. 
In QNA the Hessian matrix of second derivatives of the function to be minimized, does not need to be computed and can  
be derived by analyzing successive gradient vectors.
QNA is a generalization of the secant method to find the root of the first derivative for multidimensional problems. 
In multi-dimensions the secant equation is under-determined, and quasi-Newton methods differ in how they constrain the 
solution, typically by adding a simple low-rank update to the current estimate of the Hessian.
Since as it will be shown, this model performed the best in the GC classification problem discussed in this paper, 
we shall discuss it in more detail. 

In DAMEWARE the Quasi-Newton method has been implemented by following the known L-BFGS algorithm \citep{byrd94}. 
The QNA is an optimization of Newton based learning rule, also because, as described below, the implementation is 
based on a statistical approximation of the Hessian by a cyclic gradient calculation, that is at the base of Back 
Propagation method.
By using a local square approximation of the error function, we can obtain an expression for the minimum position. 
The gradient in every point $w$ is in fact given by:
$$ g = \nabla E = H \times \left( w - w^* \right) $$
where $w^*$ corresponds to the minimum of the error function, which satisfies the condition:
$$w^* = w - H^{-1} \times g $$

\noindent The vector $-H^{-1} g$ is known as Newton direction and it is the base for a variety of optimization 
strategies, such as the Quasi Newton Algorithm (QNA) which instead of calculating the $H$ matrix and then its 
inverse, uses a series of intermediate steps of lower computational cost to generate a sequence of matrices 
which are more and more accurate approximations of $H^{-1}$, 
These matrices are computed using only information related to the first derivative of the error function.
 
The Newton direction can be used in a line search (optimization problem) method when the Hessian matrix $H$ is 
positive definite, because under such requirement it is a descent direction. 
When the Hessian is not positive definite, the Newton direction may not be defined, because its inverse matrix 
may not exist. 
But, in addition, also when it is definite, it may not satisfy the descent trend. 
In particular, the main drawback of the Newton direction is the need for the exact Hessian matrix formulation,
which is described in more detail in Appendix \ref{AppI}.

As a matter of fact, this method was designed to optimize the functions of a number of arguments (hundreds
to thousands), because in this case it is worth having an increased iteration number due to the lower approximation precision because the overheads become much lower. This is particularly useful in astrophysical data mining problems, where usually the parameter space is dimensionally huge and is often afflicted with a low signal-to-noise ratio.

\begin{figure*}
\includegraphics[width=8.5cm]{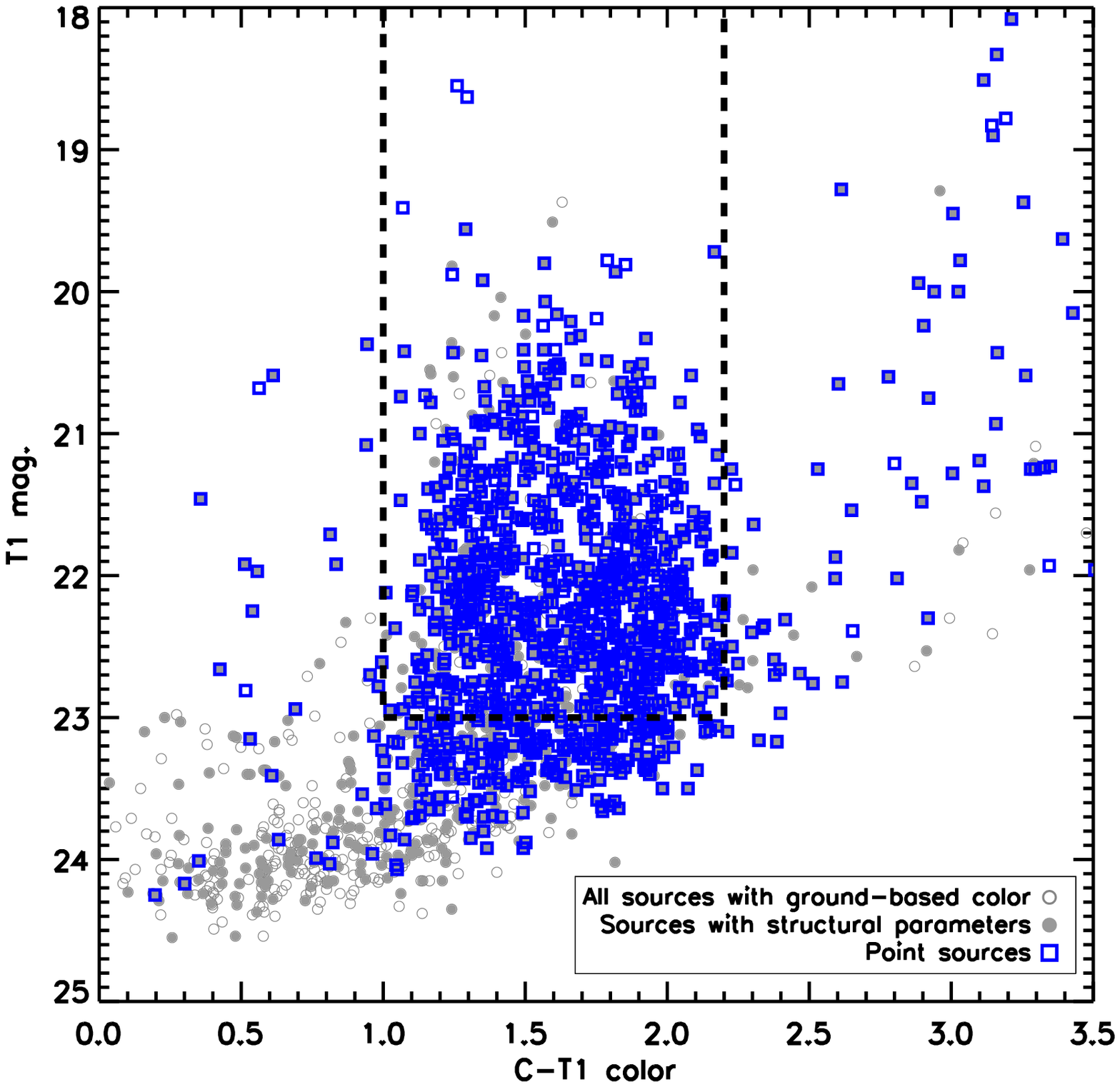}
\includegraphics[width=8.5cm]{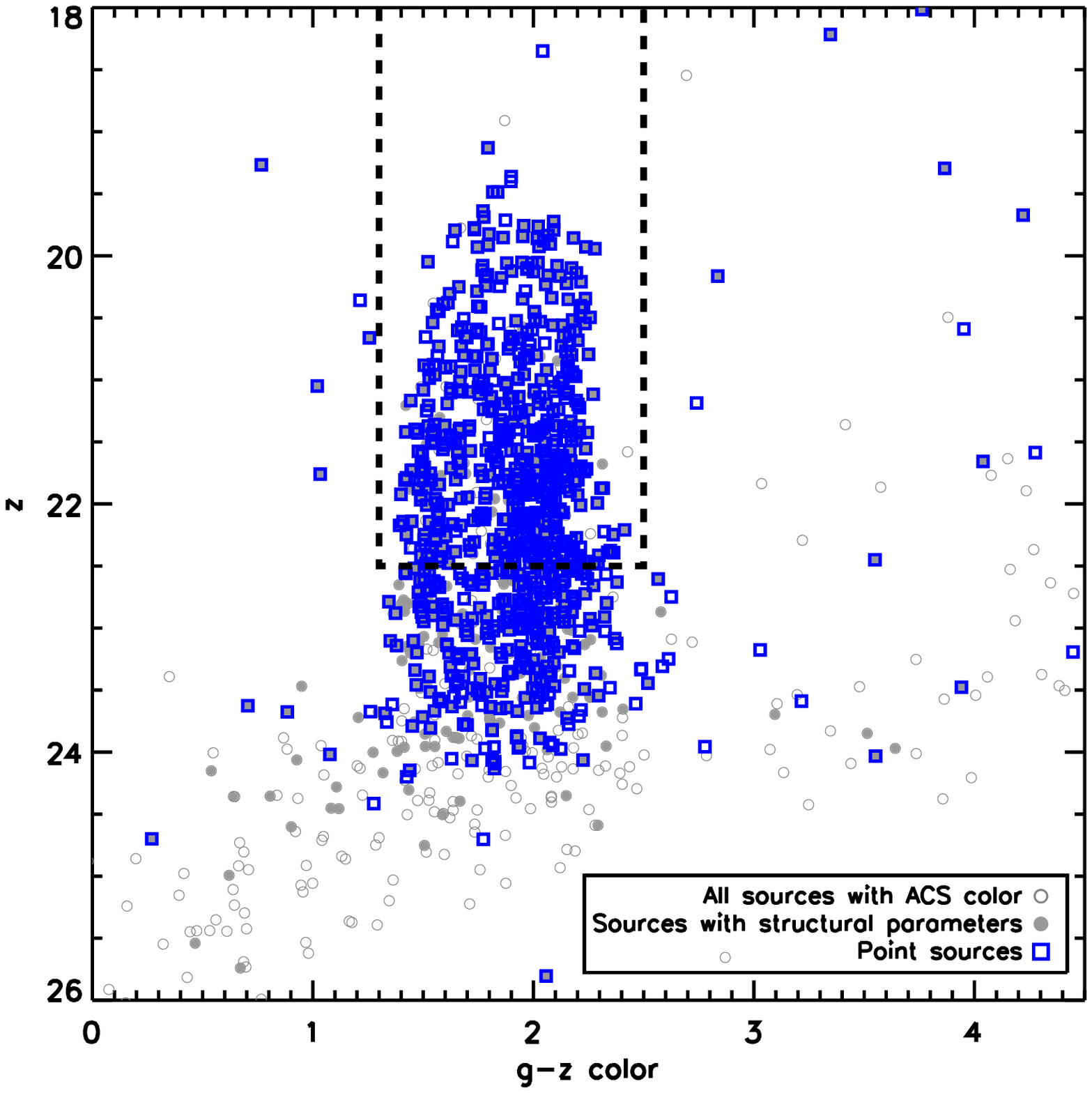}
\caption{Color-magnitude diagrams using $C\!-\!T1$ ground-based (left panel) and $g\!-\!z$ HST photometry (right panel). Ground-based photometry covers the whole FOV of our ACS mosaic, while HST colors are limited to the central ACS field ($\sim\!200\arcsec\!\times\!200\arcsec$, Figure \ref{FOV}). Open grey dots represent all sources in color catalogs while solid ones refer to the subsample with both color and structural parameters that represents our Knowledge Base. Blue squares mark pointlike sources, i.e. sources with stellarity index $>0.9$, while the dashed line highlights the parameter space (Table \ref{col_sel}) used to select bona-fide GC.}
\label{col_mag}
\end{figure*}

\section{Results}\label{exp}

\begin{table*}
\begin{tabular}{l|lrccccc}
\hline
type of experiment &	 missing features & figure of merit &	MLPQNA & 	GAME & 	SVM  &	MLPBP & 	MLPGA \\ 
\hline
complete patterns  & --	 \\
&                   & \textit{class.accuracy}		    &    98.3 &	82.1 &	90.5 &	59.9 &	66.2\\
 &                  & \textit{completeness}		     &    97.8 &  73.3 &  89.1 &  54.1 &  61.4\\
  &                 & \textit{contamination}		      &    1.8 &  18.7 &   7.7 &  42.2 &  35.1\\
\hline
no par. 11         & 11 \\
&                 & \textit{class.accuracy}		    &    98.0 &	81.9 &	90.5 &	59.0 &	62.4\\
  &                 & \textit{completeness}		      &   97.6 &  79.3 &  88.9 &  56.1 &  62.2\\
   &                & \textit{contamination}		       &    1.6 &  19.6 &   7.9 &  43.1 &  38.8\\
\hline
only optical &   8, 9, 10, 11      \\
& 	    & \textit{class.accuracy}		       &    93.9 &  86.4 &  90.9 &  70.3 &  76.2\\
  &                 & \textit{completeness}		      &    91.4 &  78.9 &  88.7 &  54.0 &  65.1\\
   &                & \textit{contamination}		       &     5.9 &  13.9 &	 8.0 &  33.2 &  24.6\\
\hline                          
mixed      &  5, 8, 9, 10, 11        \\
& 	    & \textit{class.accuracy}		     &    94.7 &  86.7 &  89.1 &	68.6 &  71.5\\
  &                 & \textit{completeness}		      &    92.3 &  81.5 &  88.6 &  52.8 &  63.8\\
   &                & \textit{contamination}		       &     5.0 &  16.6 &   8.1 &  37.6 &  30.1\\
\hline
\end{tabular}
\caption{Summary of the performances (in percentage) of the five classifiers. 
For each entry the first line refers to the classification accuracy,
while the second and third refer to completeness and contamination, respectively.}
\label{comp}
\end{table*}

As discussed in the Introduction, the purpose of this work was to implement an alternative, DM based, method to
select globular clusters in single band HST images, thus saving the observing time needed to obtain complete sets 
of multiband data.
In this section we shortly summarize the results of the series of (numerical) ''experiments'' which were performed to determine the best model and the best combination of features, while in next section 
we discuss the overall properties of the sample obtained with the DM algorithms, in comparison with traditional selection methods.

Terms like completeness, contamination, accuracy etc. are differently defined by astronomers and "data miners". 
In what follows we use the following definitions. Classification accuracy: fraction of patterns (objects) which are correctly classified (either GCs or non-GCs) with respect 
to the total number of objects in the sample; 
completeness: fraction of GCs which are correctly classified as such; 
contamination: fraction of non-GC objects which are erroneously classified as GCs. 
 
All experiments were performed on the KB sample presented in \S \ref{thedata}, assuming that bona-fide GCs are represented by
sources selected according to the color cuts in Table \ref{col_sel}. We used as features the following quantities:

\begin{itemize}
\item The isophotal magnitude (Feature 1). 	
\item Three aperture magnitudes (features 2--4) obtained through  circular apertures of radii 2, 6, and 20 arcsec respectively;
\item	The Kron radius, the ellepticity and the FWHM of the image (features 5-7); 
\item The structural parameters	(features 8-11) which are, respectively, the central surface brightness, the	core radius, the effective radius and the tidal radius. 
\end{itemize}

By making an exhaustive pruning test on all 11 dataset parameters, with the 5 machine learning models previously introduced, 
we collected a total of 425 experiments (85 per model). The details of the experiment setup can be found in Appendix \ref{AppII}.

Table \ref{comp} summarizes the most relevant results: in terms of classification accuracy and completeness, the best results (98.3\% and 97.8\% respectively) are obtained by MLPQNA using all parameters; using all available features but the number 11 (the tidal radius) we obtain marginally worse results, as can be expected given the high noise present in this last parameter, which is affected by the large background due to the host galaxy light. In terms of contamination comparable results ($\lesssim 2\%$) are obtained with the same model both with or without feature 11. We point out that since the experiment without feature 11 provides results comparable to the one using all features, but requires less information and is less computationally demanding, we consider the latter to be the case providing the highest overall performance, as usually done in DM experiments. In other words the experiment without feature 11 represents the best compromise between required overall performance and complexity of the KB.

The best result obtained without using the structural parameters is 93.9\% (classification accuracy) thus indicating that  
the availability of detailed structural parameters does indeed help to improve the results, but only by $ \sim 5\%$.
Moreover, the pruning in the mixed cases (by excluding some structural and optical features) revealed a similar 
behavior in all models, in terms of quantity of correlated information introduced by individual features in the patterns.
Five optical features (namely the isophotal and aperture magnitudes and the FWHM of the image) were recognized as the most relevant by 
all models. 
Among the structural parameters, the central surface brightness and the core radius were recognized as relevant by all models but the SVM and MLPGA models. 
In all other cases, other residual optical and structural parameters were evaluated low carriers of correlated information.

\begin{figure*}
\includegraphics[width=8.cm]{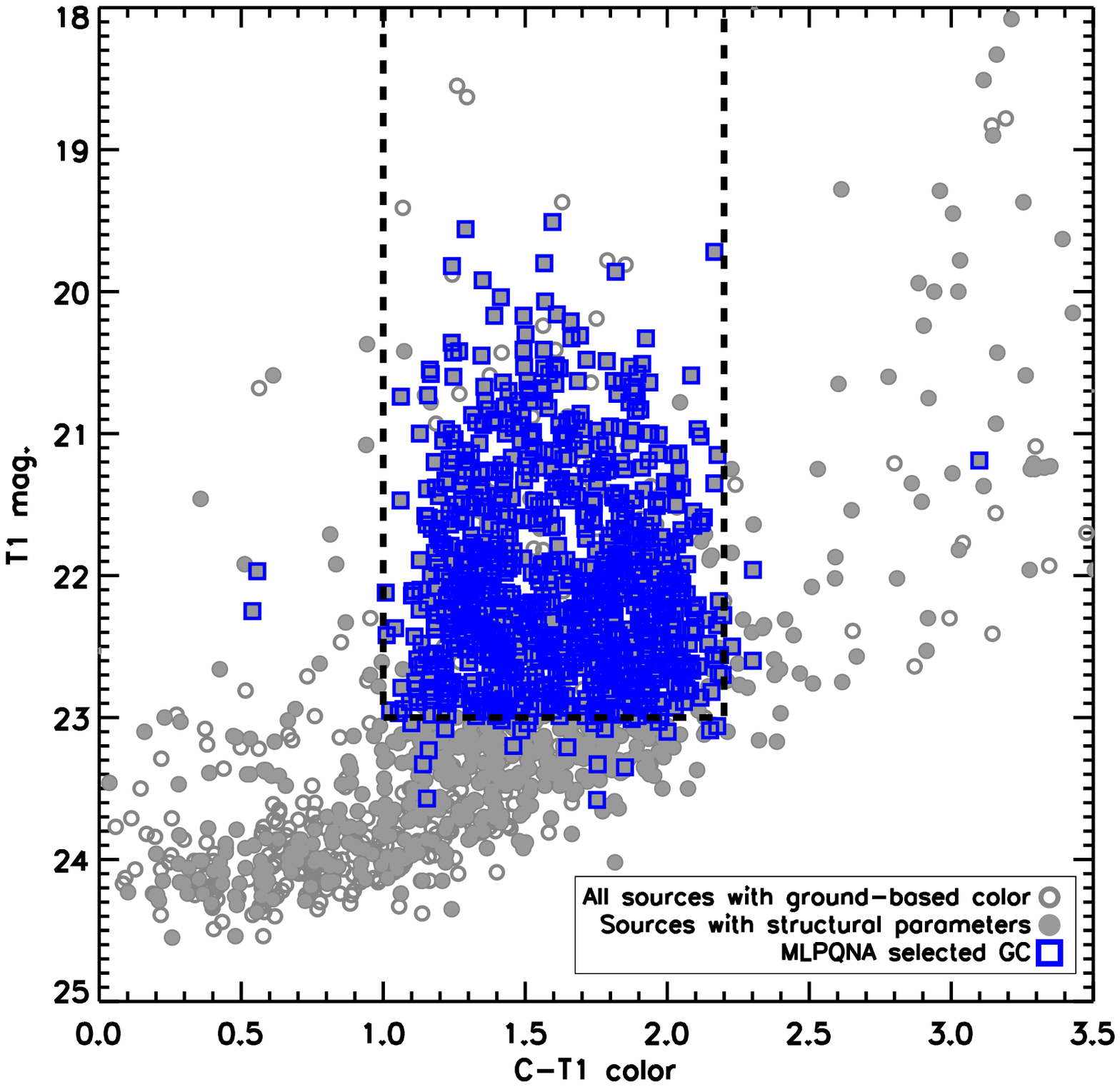}
\includegraphics[width=8.cm]{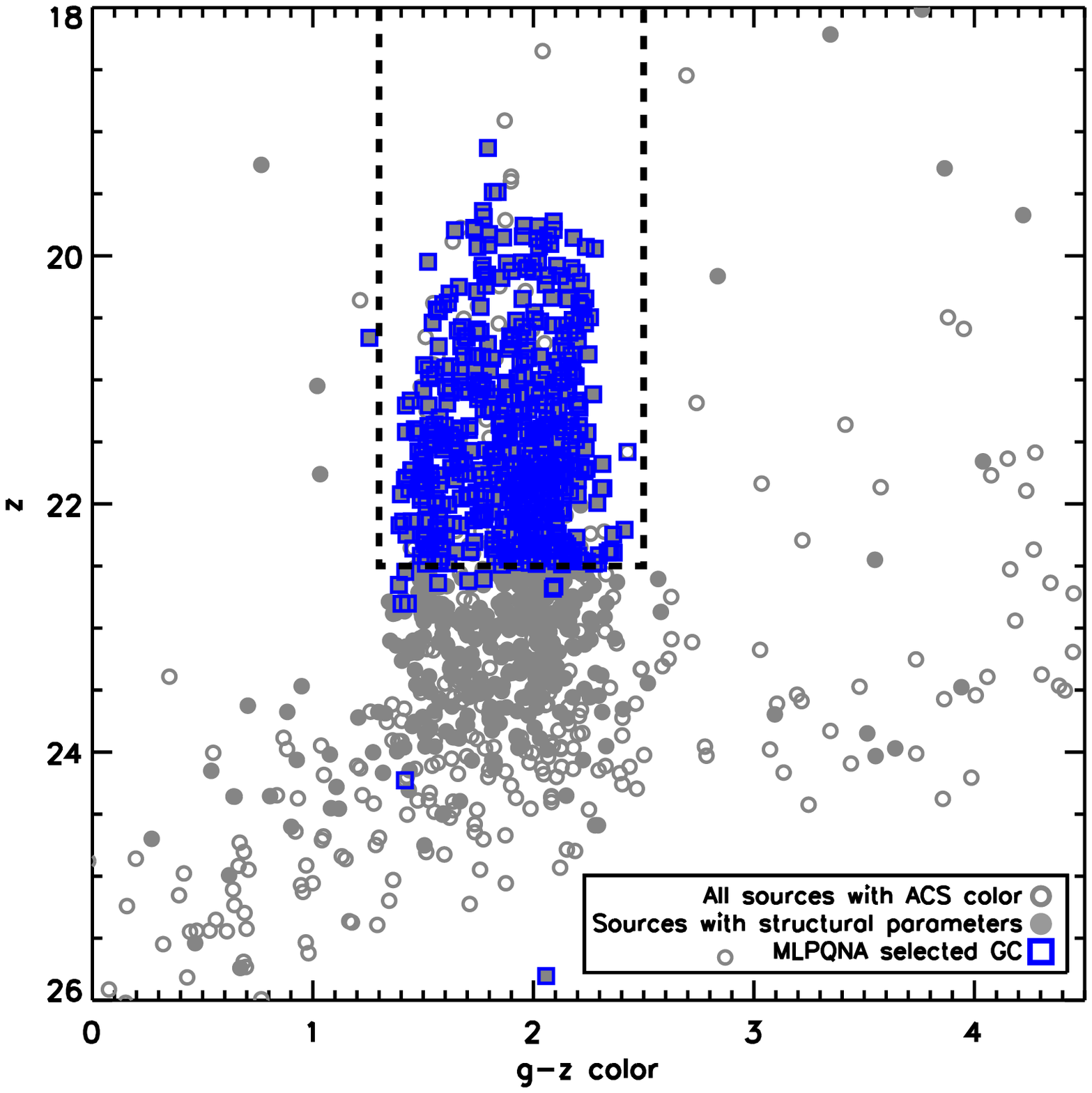}
\caption{Same as Figure \ref{col_mag} showing the color distribution of the MLPQNA selected sample. The MLPQNA sample (blue squares) reproduces the properties of the color-selected GC population (i.e. the KB) with much less contaminants than, e.g., the pointlike population shown in Figure 
\ref{col_mag}.} \label{col_mag_NN}
\end{figure*}

It is worth pointing out that the performances, in terms of completeness and contamination, quoted above are all derived with respect to the test sample (and thus ultimately from the KB), and do not include possible biases affecting the KB itself. Such biases will be propagated to the final sample by any DM algorithm, as these rely on the assumption that the KB is a fair and complete representation the ``real'' population that we want to identify. Thus if the KB is severely incomplete or contaminated, this is a separate issue that has to be addressed in the training sample selection phase.

\section{discussion}\label{disc}
In order to test the effectiveness of our method, we need to compare its performances with those 
offered by more traditional approaches. For homogeneity (same data set) we shall use as template 
the method discussed in \citet{paolillo11} which used a selection criteria based on magnitude 
and morphology. 
Figure \ref{maghisto} shows that sources with \textsc{SExtractor} stellarity index $>0.9$ (grey solid line) 
are distributed as the GC luminosity function down to $m_V=26$, while at fainter magnitudes background unresolved 
sources dominate the overall sample. 
 Based on these considerations  Paolillo et al. choose as GC candidates sources having stellarity index $>0.9$ and $m_V<26$ mag. 
Clearly a more sophisticated selection process, based on complex combinations of photometric and structural 
parameters \citep[see for instance][]{puzia12}, could be adopted but any such approach requires anyway extensive 
testing to verify what biases are introduced in the final sample and it is not clear how such biases can be evaluated and 
corrected for without the availability of additional data (e.g. more uniform color coverage or random background fields to compare with).

From Figure  \ref{col_mag} it can be seen that although the use of the stellarity and magnitude criteria effectively selects the bulk of the color-selected GC population, there are sources consistent with GC colors, which are missed by this approach; on the other hand this subsample includes many objects outside the allowed color range.  
We can calculate the level of completeness and contamination resulting from the 
simple approach of \citet{paolillo11}, as done in \S \ref{exp} for the DM methods. We derive two different estimates: i) for the central region covered by the more accurate $g$ and $z$ HST photometry and,
ii) for the entire field covered by the ground based $C$ and $T1$ data.
Within the central region $92\%$ of our GC candidates (within $m_V<26$ by definition) are consistent with the $1.3\leq g-z<2.5$ 
color cut and $z<22.5$. Using the $C-T1$ photometry instead, which extends over the whole HST mosaic we find that $82\%$ of the GC candidates are consistent with the $1.0\leq C\!-\!T1<2.2$ color and $T1<23$ magnitude cuts.
On the other hand, $\sim 4\%$ and $\sim 9\%$ of the GC candidates have respectively $g-z$ and $C-T1$ colors outside the allowed range as given in Table \ref{col_sel}. 

When these numbers are compared with those presented in Table 1, we see that the MLPQNA outperforms the simpler approach 
used by \citet{paolillo11} both in the central region and across the whole field, in the sense that it results in higher completeness, retrieving a larger fraction of the color-selected sources using only single band photometry. 
GAME and SVM may still perform better in the galaxy outskirts, although in the galaxy center they are slightly less accurate.
In terms of contamination the MLPQNA again performs better than the \citet{paolillo11} approach, yielding $<2\%$ spurious sources in the two best experiments (\textit{complete patterns} and \textit{no par.11}). The other MLPQNA experiments and all SVM cases are still competitive in the galaxy outskirts.

\begin{figure}
\includegraphics[width=8.5cm]{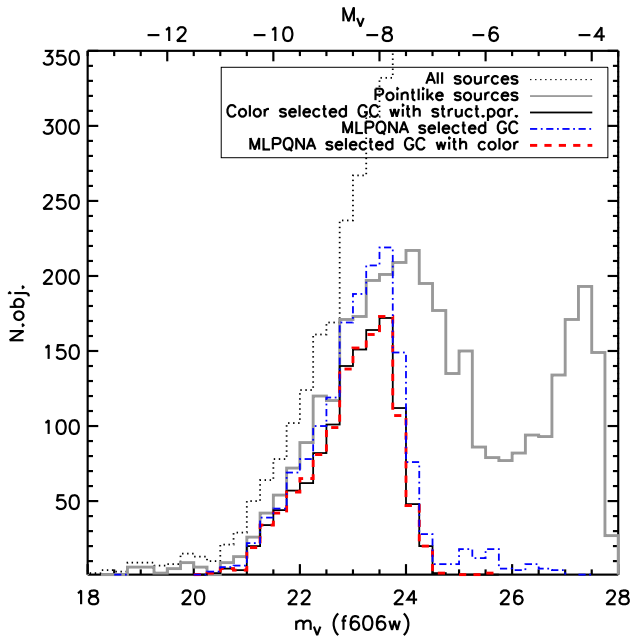}
\caption{Same as Figure \ref{maghisto} but for the MLPQNA selected samples. 
The MLPQNA approach (dashed red line) is able to retrieve almost the entirety of the color-selected GC population (solid black line);
applying the same algorithm to all sources with structural parameters (but no color, blue dot-dashed line) we can thus retrieve many more objects than available in the color-selected subsample, sharing the same luminosity distribution of the latter population.}
\label{maghisto_NN} 
\end{figure}

The performance of the MLPQNA method is better understood looking at the color-magnitude plot shown in Fig. \ref{col_mag_NN}. 
The MLPQNA sample reproduces the properties of the color-selected GC population with much less contaminants than, e.g., 
the pointlike population shown in Fig. \ref{col_mag}, and less outliers.
In Fig. \ref{maghisto_NN} we show the luminosity distribution of the MLPQNA sample:
the MLPQNA approach (dashed red line) is able to retrieve almost the entirety of the color-selected GC population (solid black line). We point out that the luminosity limit at $m_V\sim 24$ is due to the magnitude threshold imposed on the color-selected sample (Table \ref{col_sel}) in order to get a uniform limit across the whole color range (Figure \ref{col_mag_NN}) and FOV, and is thus not an intrinsic feature of the GC luminosity function which extends down to $m_V \gtrsim 26$ mag. 

\begin{figure*}
\includegraphics[width=8.5cm]{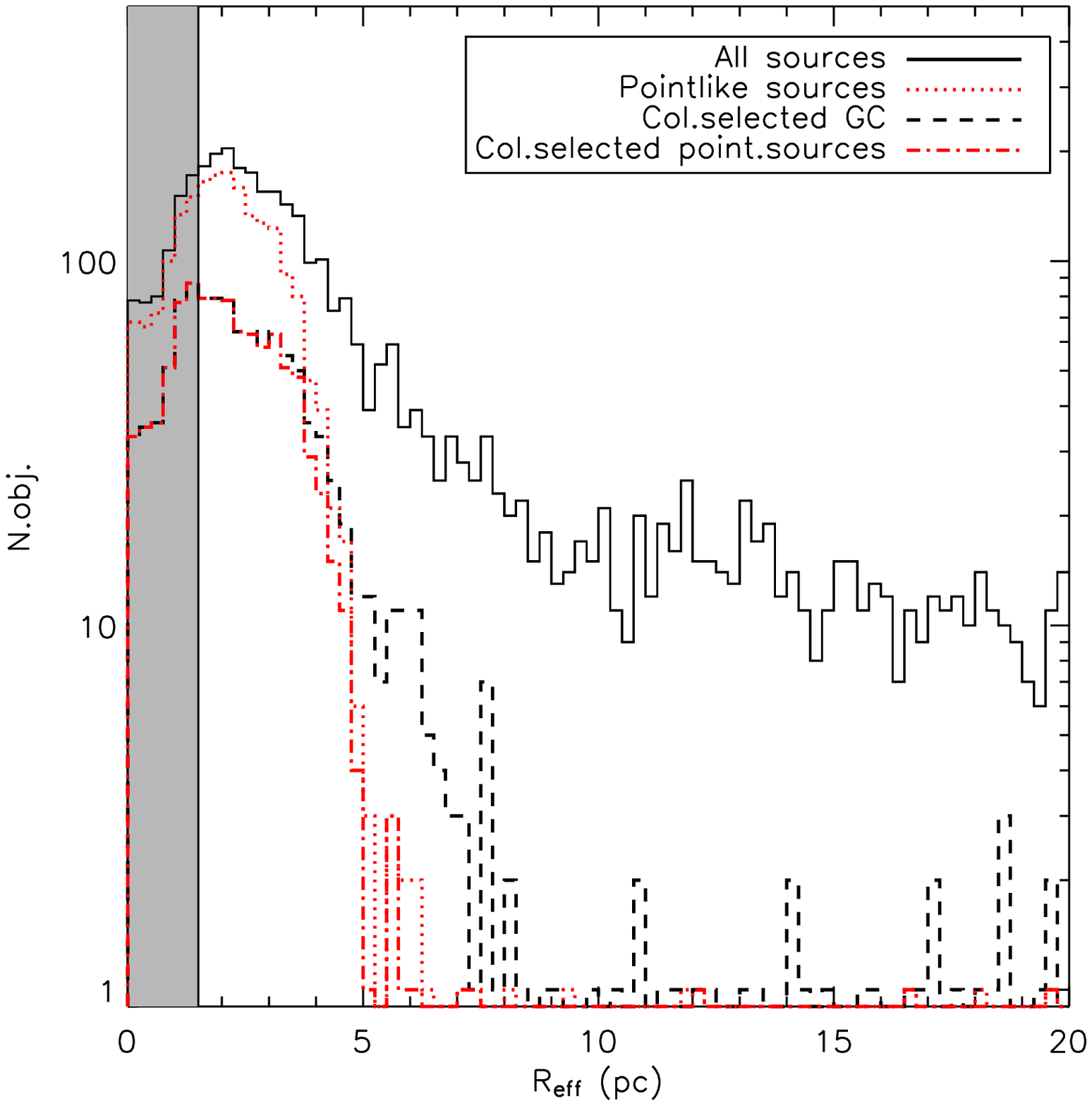}
\includegraphics[width=8.5cm]{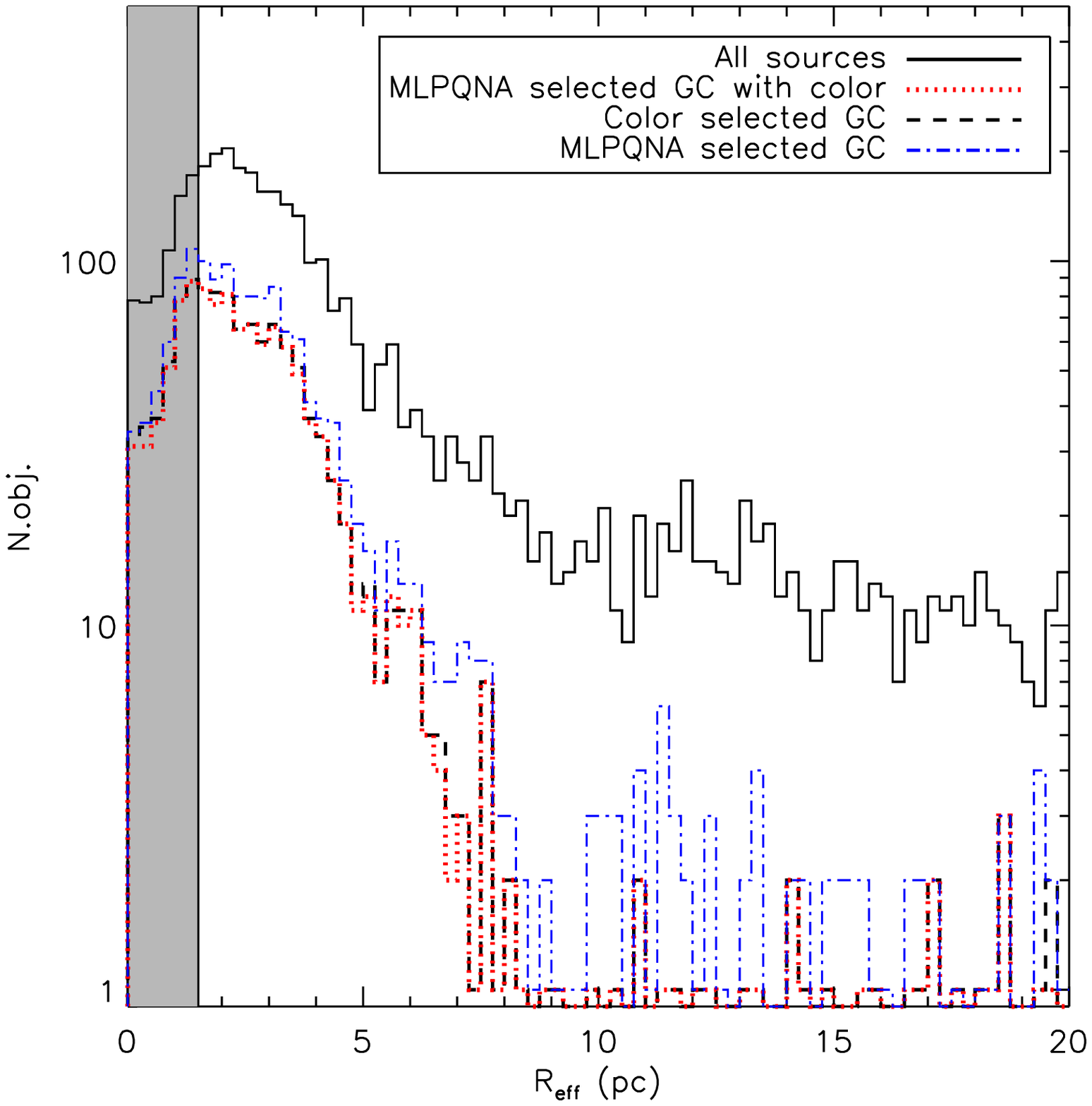}
\caption{\textit{Left panel:} Half-light radius distribution for the entire ACS optical catalog (solid line), compared to \citet{paolillo11} GC candidates, i.e. pointlike sources with $m_V<26$ (dotted line). Restricting the sample to color-confirmed GCs (dashed and dot-dashed lines) shows that the \citet{paolillo11} selection criteria misses very extended GCs with $R_{\rm eff}>5$ pc. The shaded region highlights the region where our size measurement are poorly constrained (see \citealt{paolillo11, puzia12}). \textit{Right panel:} Same as left panel but for the MLPQNA selected samples. 
The MLPQNA selected sample (dotted red line) reproduces the size distribution of the the color-selected GC population (dashed black line), thus avoiding the size biases resulting from the simpler \citet{paolillo11} selection criteria; the same is true when applying the MLPQNA algorithm to the larger subsample with structural parameters (blue dot-dashed line).}
\label{size_distro}
\end{figure*}

A detailed investigation of the properties of the spurious sources is difficult since the strength of DM algorithms is to detect \textit{hidden} correlations among the parameters, and use them to classify unknown sources; this however means that such correlations are hard to identify through a simple (and low-dimensional) view of the source distribution in the parameter space. In our specific case we found that most contaminants are indeed GCs which fail the color-magnitude classification technique by only one criteria (e.g. they lie just outside the chosen color or magnitude range, see Figure \ref{col_mag_NN}). It is thus unsurprising that the MLPQNA identifies such sources as GCs as all other parameters obey the correlations identified in the training phase. A few more extreme objects are found to be affected by photometric or structural problems in the data, such as the overlap with a nearby source which may introduce severe contamination in low-resolution ground based data, or a position close to the chip gap in HST data. A few can also be expected to be foreground stars which are misclassified due to the small angular size of some GCs in the training sample, lying at the resolution limit of HST data (gray region in Figure \ref{size_distro}).

An additional advantage in the use of DM techniques can be see comparing the structural parameters of pointlike sources 
with $m_V<26$ mag with those of the 
color-selected subsample (Figure \ref{size_distro}, left panel): we find that the \citet{paolillo11} selection criteria misses 
extended sources with $R_{eff}\gtrsim 5$ pc, as it can be expected given the compactness requirement (stellarity index $>0.9$).
The right panel of the same Figure shows that the MLPQNA methods is instead able to retrieve also the most extended GCs. 
While some of these extended sources may be background galaxies, we point out that the most extended GCs, such as the Galactic GC $\omega$-Cen, do fall in this range. In fact we used the subset of GC confirmed through radial velocity measurements \citep{Dirsch2004} to verify that a significant fraction ($\sim 10\%$) of the NGC1399 GC population has $R_{eff}\gtrsim 5$ pc, and that the size distribution of this subsample is statistically indistinguishable from both the color- and  MLPQNA-selected populations. Obviously we cannot confirm that \textit{all} extended sources are genuine GCs but, as already discussed in \S \ref{exp}, we emphasize that the performance of the method has to be evaluated only by its ability to retrieve the same sources included in the training sample, i.e. the color-selected GCs in our experiment.

Applying the same algorithm to the larger ensemble of sources with structural parameters (but no color information) we are now able to retrieve more objects than available in the color-selected subsample, sharing very similar properties to the latter population. The population of MLPQNA selected GCs identified within the whole population is shown in Figure \ref{maghisto_NN} and Figure \ref{size_distro} (right panel) as a dot-dashed line. In our specific test case (e.g. NGC1399) this method allows to identify $\sim 30\%$ more GCs than relying the subsample of sources with color; this larger sample closely follow the GC LF down to the magnitude limit imposed by the color selection, as well as the structural properties of the bona-fide GC population. Thus the gain with respect to other selection techniques is in the ability to retrieve a larger population with well defined properties, at lower observational cost. In other programs the gain can be much larger: for instance in cases of large surveys where DM algorithms can be trained on a KB consisting on a limited number of multi-band observations covering only a small fraction of the FOV; the trained algorithm will then allow to extract statistically equivalent samples from the entire survey.

Finally we note that each experiment was not really time consuming. It was executed on a common desktop multi-core PC in a multi-threading environment, resulting in about 3600 sec (1 hour) of duration for the training phase in the worst case (i.e. on the whole dataset patterns with all 11 features). The test phase is instantaneous, since the trained network acts like a one-shot function.
Of course the complexity and indeed the execution time depends in a quadratic form on the dataset dimension. But in case of small datasets, like the present one, this is not an issue.
Besides computing time, the relevant result is that the proposed MLPQNA model revealed a strong performance also in case of small datasets where, as known, machine learning method performances are usually degrading, due to the limited size of the training samples. This is demonstrated by the poorer results obtained by other methods, shown in Table \ref{comp}, which usually perform significantly better on larger datasets.

\section{Conclusion}
We performed an experiment showing that the use of Data Mining (DM) techniques on small datasets, allows to solve complex astronomical problems such as the selection of Globular Cluster candidates in external galaxies, from single band images, provided that a subsample of sources can be used to train the DM algorithm. Since such methods do not assume any a-priori model of the population we are looking for, they allow to retrieve samples which share the same properties of the training sample and are affected by less biases than results using simpler selection techniques.  

In principle we could use more refined approaches than those tested here, such as the use of radial velocity (RV) measurements to improve the the reliability of the Knowledge Base (KB), but any such approach would require the availability of additional data i.e., in this particular case, spectroscopic observations. Such type of data are difficult to obtain and expensive in terms of observing time, thus justifying the DM methods proposed in this work. Obviously in some instances these data could already be available in the archives, as for the NGC1399 case were they have been used to verify some of our results (\S \ref{disc}).

As a closing remark, we can safely state that, in the emerging scenario of the data-driven science, a Data Mining based approach to data
analysis and interpretation seems to provide a large competitive edge over classical methods in particular for what concerns the ability to recognize patterns and derive correlations in high dimensionality dataset that are not easily handled by human perception.

\subsection*{Acknowledgments}
The authors wish to thank the whole DAMEWARE working group, whose huge efforts made the DM facility available to the scientific community.
\noindent MB acknowledge support from PRIN-INAF 2010.
\noindent MP acknowledge support from PRIN-INAF 2009, and thanks the ASI Science Data Center (ASDC) for support and hospitality.
\noindent GL wishes to thank Prof G.S. Djorgovski and the whole Department of Astronomy at the California Institute 
of Technology in Pasadena, for hospitality. We also thank the anonymous referee for useful suggestions and comments.

\vspace{2cm}

\appendix
\section{Quasi-Newton learning rule}
\label{AppI}
Quasi-Newton direction search methods provide a very useful alternative in that they do not require a precise calculation of the Hessian.
In place of the Hessian matrix $H_k$, they use an approximation matrix $A_k$, updated after each iteration $k$, 
to take into account of the additional information gain obtained. 
The cyclic updates make use of the gradient changes, which at each step k provides information about the second derivative of the error function $f_k$ along the optimization search direction.
More rigorously, given $x_k$ a partial solution to the optimization problem at the iteration $k$ 
(we want to converge to the optimal solution $x^*$), when $x_k$ and $x_{k+1}$ lie near the optimal solution $x^*$, 
within which $H(x)$ is positive definite, we can write:
$$H_k \left(x_{k+1}-x_k \right) = \nabla^2 f_k\left(x_{k+1}-x_k \right) \sim \nabla f_{k+1} - \nabla f_k$$
The Quasi-Newton method chooses the Hessian approximation $A_{k+1}$ so that it can well represent the true Hessian. 
In other words we require to follow the well-known secant equation condition:
$$A_{k+1} \left(x_{k+1}-x_k \right) = \nabla f_{k+1} - \nabla f_k$$

\noindent For completeness we recall also that the previous equation is intrinsically defined under additional 
conditions, such as symmetry (typically assumed by the exact Hessian) and the low rank of the difference between 
successive approximations $A_k$ and $A_{k+1}$.
In the MLPQNA model we apply the Hessian approximation known as the BFGS formula, named after its discoverers 
\citep{broyden70, fletcher70, goldfarb70, shanno70}. 
This is defined by the following equation:
Let us call $sol_k= x_{(k+1)}-x_k$ and $g_k=\nabla f_{(k+1)}- \nabla f_k$, the respective matrix terms of the Eq. 2 
we obtain the following rank-two matrix

$$A_{k+1} = A_k - \frac{A_k \ sol^T_k A_k}{sol^T_k A_k sol_k} +\frac{g_k g^T_k}{g^T_k sol_k}$$

The BFGS formula generates positive definite approximation matrices under the condition that the initial approximation matrix 
$A_0$ is positive definite and the term $g^T_k sol_k > 0$ L-BFGS. 
From a computational point of view, the BFGS formula is time-consuming and requires storing at each step a dense $N\times N$ 
approximation matrix. Dealing with massive data optimization problems, in order to overcome such requirements, we decided to implement a limited-memory algorithm, known as L-BFGS \citep{zhu97}.

\noindent The L-BFGS stores at each step only few vectors of length n that represent the approximations implicitly. Despite this improvement in the storage requirements, it yields an acceptable (almost linear) rate of convergence. The main idea of this method is to use error function curvature information from only the most recent iterations to construct the Hessian approximation.
Of course, the final result will not be the Hessian itself but just an approximation. Surprisingly enough, while the convergence slows down, performances are not affected much and may even improve since it depends on the number of processor's time units spent to calculate the result.

\section{Setup of the experiments}
\label{AppII}

\noindent In the following sections the feature are referred to the cardinal number (feature 1: MAG\_ISO, etc).
For each model we choose the configuration parameters in order to perform the best results.\\

\subsection{Multi Layer Perceptron trained by Back Propagation (MLP-BP)}
\begin{itemize}
\item Input Nodes (equivalent to the number of features considered in the dataset patterns)		
max number: 11	(complete patterns);	min number: 4 (pruning on optical features);nominal number: 7 (complete optical dataset);	

\item Hidden Nodes (depending on the number of features considered in the dataset patterns). 
Max number: 23	(with input nodes in [8, 11]);	min number: 15 (with input nodes in [4, 7]);	

\item Output Nodes (based on crispy classification): 2 (1 0 GC, 0 1 not GC);
	
\item Activation Functions (neuron function type, used to provide its output, by processing inputs).
input layer: (no input processing, just propagate it);		
hidden layer: nonlinear hyperbolic tangent of input;		
output layer: linear with softmax normalization  (outputs  sums  up  to  1.0  and  converge to posterior probabilities);		

\item Learning Rule Parameters. Output Error Type: Cross Entropy;	Training Mode: Batch (weights update after each whole dataset patterns calculation);	Training Rule: Back Propagation with Conjugated Descent Gradient;	Error Loop Threshold: 0.001 (one of the stopping criteria);	Number of Iterations: 10000 (one of the stopping criteria);
\end{itemize}
			
\subsection{Support Vector Machines (SVM)}
\begin{itemize}
\item Model: C-Support Vector Classification (C-SVC); Kernel: Radial Basis Function;

\item Gamma (for each experiment we have a multiplicative step). Min number: $2^{-15}$; max number: $2^{23}$; 
step: $4$(multiplicative). $C$ (for each experiment we have a multiplicative step). Min number: $2^{-5}$; 
max number: $2^{15}$; step: $4$ (multiplicative);

\item Error Tolerance: 0,001;

\item Cache: 100MB;

\item Shrinking: On;

\item Probability Estimates: Off;

\item Cross Validation: k-fold (k = 5);

\item Weights: 1;
\end{itemize}

\subsection{Genetic Algorithm Model Experiment (GAME)}
\begin{itemize}
\item Model: Genetic algorithm with fitness based on trigonometric polynomial expansion;

\item Topology: population of chromosomes, each of them composed by genes;	

\item Input features (depending on the number of features considered in the dataset patterns). 
Max number: 11 (complete dataset); min number: 4 (pruning on optical features); nominal number: 7 (complete optical dataset);

\item Genetic Population Size (depending on the number of features and polynomial order). 
Max number: 67 (with 11 features); min number: 25 (with 4 features);

\item population size: $(polynomial_{order} * num_{features}) + 1$;

\item Genetic Chromosome Size (depending on the polynomial order). Number: 13 (with polynomial order = 6);
chromosome size: $(2 \times polynomial_{order}) + 1$;

\item Output (based on crispy classification). Number in BoK: 1 (0 if no GC; 1 else);

\item Output Error Type: TMSE (Thresholded Mean Square Error) with threshold 0,4;

\item Error Loop Threshold: 0,001 (one of the stopping criteria);

\item Polynomial Order: 6;

\item Tournament Selection (based on the Wheel Roulette, max probability on the entire population fitness). 
Number of Tournament Chromosomes: 2;

\item Genetic Operators. Crossover Probability: 0,9; Mutation Probability: 0,2; Elitism Factor: 2;

\item Initial Population Distribution: gaussian standard, with all values generated into range [-1, +1];

\item Number of Iterations: 10000 (one of the stopping criteria);
\end{itemize}

\subsection{Multi Layer Perceptron trained by Quasi Newton (MLPQNA)}
\begin{itemize}
\item Input Nodes	(depending on the number of features considered in the dataset patterns). Max number: 11 (complete dataset);
min number: 4 (pruning on optical features);
nominal number: 7 (complete optical dataset);

\item Hidden Nodes (depending on the number of features considered in the dataset patterns). Max number: 23 (with input nodes in 
[8, 11]);	min number: 15 (with input nodes in [4, 7]	);
\item Output Nodes (based on crispy classification): number in BoK: 1 (0 if no GC; 1 else);
\item Activation Functions (neuron function type used to provide its output, by processing inputs). Input layer: no input processing, just propagate it; hidden layer: not linear hyperbolic tangent of input; output layer: linear with softmax normalization  (outputs  sums  up  to  1.0  and  converge to posterior probabilities).
\end{itemize}

\noindent Learning Rule Parameters
\begin{itemize}
\item Output Error Type: Cross Entropy;
\item Training Mode: Batch (weights update after each whole dataset patterns calculation);
\item	Training Rule: Quasi Newton (inverse hessian approximation by error function gradients);
\item QNA Implementation Rule: based on L-BCFG method (L is for Limited memory);
\item QNA Parameters. Decay: 0,001 (weight decay during gradient approximation);
Restarts: 20 (random restarts for each Approximation Step);
Wstep: 0,01 (stopping threshold, min error for each Step);
MaxIts: 1500 (max number of  Iterations for each Approx. Step);
\end{itemize}

\subsection{Multi Layer Perceptron trained by Genetic Algorithms (MLPGA)}
\begin{itemize}
\item Input Nodes (depending on the number of features considered in the dataset patterns).
Max number: 11 (complete dataset);	
min number: 4 (pruning on optical features	);
nominal number: 7 (complete optical dataset);
\item Hidden Nodes (depending on the number of features considered in the dataset patterns)
max number: 23 (with input nodes in [8, 11]);	
min number: 15 (with input nodes in [4, 7]);	
\item Output Nodes (based on crispy classification).
number in BoK: 1 (0 if no GC; 1 else);	
\item Activation Functions (neuron function type used to provide its output, by processing inputs).
Input layer: no input processing, just propagate it;		
hidden layer: nonlinear hyperbolic tangent of input;		
output layer: nonlinear hyperbolic tangent of input;		
\item Learning Rule Parameters.
Output Error Type: MSE;			
Training Mode: Batch (weights update after each whole dataset patterns calculation);	
Training Rule: Genetic Algorithm with Roulette Wheel selection function and fitness based on the MSE between target and output of dataset patterns;
\item MLPGA Parameters.		
Genetic Population Size: 25;
Genetic Chromosome Size: 13;
Error Loop Threshold: 0,001;
Tournament Selection: based on the Wheel Roulette method (max probability on the entire population fitness);	
Number of Tournament Chromosomes: 2;
Crossover Probability: 0,9;
Mutation Probability: 0,2;
Elitism Factor: 2;
Initial Population Distribution: gaussian standard, with all values generated into range [-1, +1];
Number of Iterations: 10000 (one of the stopping criteria).
\end{itemize}

\label{lastpage}
\end{document}